\documentclass{customized}

\usepackage{graphicx}
\usepackage[percent]{overpic}
\usepackage{epstopdf, epsfig}
\usepackage{graphicx}
\usepackage{dcolumn}
\usepackage{bm}
\usepackage{color}
\usepackage{array}
\usepackage{amsmath,amssymb,stmaryrd} 
\usepackage{multirow}
\usepackage{booktabs,comment}

\definecolor{blue3}{rgb}{0, 0.1770, 0.3410}

\begin{document}  

\newtheorem{lemma}{Lemma}
\newtheorem{corollary}{Corollary}

\shorttitle{End boundary effects on wakes of inclined cylinders} 
\shortauthor{K. Zhang \textit{et al.}} 

\title{End boundary effects on wakes dynamics of inclined circular cylinders}

\author
 {
 Kai Zhang\aff{1,2}\corresp{kai.zhang@sjtu.edu.cn}
, Yan Bao\aff{1}\corresp{\email{ybao@sjtu.edu.cn}}, Dai Zhou\aff{1} \and
Zhaolong Han\aff{1}
}

\affiliation
{
\aff{1}State Key Laboratory of Ocean Engineering, School of Naval Architecture, Ocean and Civil Engineering, Shanghai Jiao Tong University, Shanghai 200240, China
\aff{2}Department of Mechanical and Aerospace Engineering, Rutgers University, Piscataway, NJ 08854, USA
}

\maketitle

\begin{abstract}

We perform direct numerical simulations to characterize the three-dimensional wake dynamics of long inclined circular cylinders with inhomogeneous end boundary conditions.
Three Reynolds numbers, $\Rey=100$, 200 and 300,  corresponding to the regimes of laminar flow, mode A* wake, and mode B wake, respectively, are considered to reveal the roles of the intrinsic secondary instabilities and the extrinsic end boundary effects in shaping the three-dimensional flows.
At $\Rey=100$, the end boundary effects are felt over the entire cylinder span by inducing oblique vortex shedding, which is associated with stronger spanwise flow in the wake than a parallel shedding.
The Strouhal number of the oblique shedding is related to that of the parallel shedding of straight cylinder by the cosine law, considering the combined inclination angle and oblique angle.
At $\Rey=200$, the intrinsic secondary instability results in large-scale vortex dislocation, precluding the propagation of the end boundary effects towards further span.
Nevertheless, for the inclined cases, oblique shedding is still observed within limited span from the upstream end boundary.
The oblique vortices feature intact and straight vortex cores, and are related to the two-dimensional flow at $\Rey=200$ (that are void of vortex dislocations) from the viewpoint of cosine law.
Further along the span, the oblique vortices destabilizes with the formation of small-scale vortices, and the flow transitions to the typical mode A* wake.
At $\Rey=300$, the highly three-dimensional flow near the end boundary creates disturbances that travel along the cylinder span, creating vortex dislocations for cases with low inclination angles.
For high inclination angle, oblique vortex shedding is again observed over the cylinder span, and is not disrupted by vortex dislocations of either intrinsic or extrinsic causes.
The present study offers renewed insights into the three-dimensional wake dynamics of inclined cylinders, and lays the foundation for the designs of long flexible engineering structures and related flow control techniques.

\end{abstract}

\section{Introduction}
\label{sec:intro}
Flow over a circular cylinder has received sustained attention due to its fundamental and practical importance.
As the key feature of the cylinder wake, the K\'arm\'an vortex street is characterized by periodic shedding of counter-rotating vortices from the cylinder, and is related with various phenomena such as the unsteady hydrodynamic forces, flow-induced vibrations, noise, etc.
Over the past few decades, substantial studies have been carried out to understand various aspects of cylinder wake \citep{williamson1996ARFM,zdravkovich1997flow1,zdravkovich1997flow2,norberg2001flow,jiang2016three,cheng2017large,garcia2019span,fan2019mapping,mittal2021cellular}.
In particular, the three dimensionality in the wake of the cylinder has been the focus of extensive research.

Three-dimensional wake can arise from both ``intrinsic'' and ``extrinsic'' mechanisms.
For an infinitely long circular cylinder that is void of extrinsic boundary effects, the three-dimensional flow emerges as the Reynolds number increases beyond $\Rey\sim 188$ \citep{barkley1996three,henderson1996secondary}.
This intrinsically three-dimensional flow forms via the secondary instabilities of the periodic vortex street, and manifests itself successively as the mode A, mode A*, and mode B wakes, with increasing Reynolds number.
The mode A wake is characterized by wavy spanwise vortex cores interlaced with streamwise vortex pairs, with a wavelength of about 4 diameters.
The formation of mode A is shown to be originated from an elliptic instability of the primary vortex cores, and the streamwise vortex pairs are formed through Biot-Savart induction \citep{williamson1996JFM,leweke1998three,thompson2001physical}.
The pure mode A wake occurs in the beginning phase of the three-dimensional transition, and only lasts for a short period of time before it evolves into the mode A* featuring large-scale vortex dislocation.
Further increasing the Reynolds number to $\sim 230$, the mode B wake characterized by fine-scale streamwise vortices (wavelength of $\sim$1 diameter), but more stable spanwise vortex cores, becomes the main form of three dimensionality in the cylinder wake.
Unlike the mode A, the streamwise vortices in mode B are formed in the braid shear layer region due to a hyperbolic instability \citep{williamson1996JFM,leweke1998three,thompson2001physical}.
The transitions of these wake modes are associated with discontinuities in the shedding frequencies of the K\'arm\'an vortex streets with respect to the Reynolds number.

In reality, the wake dynamics of cylinders are inevitably influenced by the end boundary conditions, which serve as ``extrinsic'' mechanism for the emergence of three-dimensional flows.
The flow structures near the inhomogeneous boundary are highly three-dimensional due to the formation of tip vortices in the case of a free end \citep{park2000free,roh2003vortical,krajnovic2011flow,sumner2013flow,cao2022topological}, and the interaction with boundary layer flow in the case of a nonslip wall \citep{slaouti1981experimental,pattenden2005measurements,krajnovic2011flow,mittal2021cellular}.
At low Reynolds numbers, such complex three-dimensional flows typically occupy 10 -- 20 diameters near the end boundary \citep{williamson1989oblique}. 
The oblique vortex shedding are observed for the rest of the cylinder span, even hundreds of diameters in length. 
Depending on the boundary condition, the Reynolds number and the aspect ratio of the cylinder, the vortex shedding can be split into different cells with distinct shedding frequencies, separated by vortex dislocation in between \citep{behara2010flow,mittal2021cellular}.
Taking advantage of the high sensitivity of the cylinder wake to the end conditions, \citet{williamson1989oblique} achieved both oblique shedding and parallel shedding behind a cylinder by manipulating the end plates.
This flow control technique has also been reported to be effective for higher Reynolds number flows  \citep{prasad1997JFM,luo2005parallel}.

The wake of an inclined cylinder presents another type of inherently three-dimensional flow, due to the strong axial velocity generated behind the cylinder \citep{van1968experiments,matsumoto1990aerodynamic,najafi2016time,zhang2018large}.
In many cases, the wake is featured by shedding of vortex tubes that are parallel with the cylinder axis, and the spanwise flow component can be decoupled from the rest two dimensions. 
This leads to the famous independence principle (also known as cosine law), which states that the aerodynamic characteristics of an inclined cylinder are related to those of a straight cylinder with a factor of $\cos\beta$, where $\beta$ is the inclination angle.
The validity of independence principle has been theoretically proven by \citet{sears1948boundary} for boundary layer flows over infinite yaw cylinders.
For the separated wakes over circular cylinders, the independence principle is shown to hold true for cases with inclination angle smaller than $\sim 45^{\circ}$ \citep{lucor2003effects,zhao2013three,bourguet2015validity}.
However, most studies on inclined cylinders (particularly in computational studies) have employed the assumption of spanwise homogeneity, and have not considered the effects of end boundary condition on the flow.

The wake dynamics of an inclined cylinder with inhomogeneous end conditions can exhibit complex features.
For short inclined cylinders, the flows are dominated by vortical structures that develop near the ends.
\citet{choi2007endwall} experimentally studied the heat transfer around the circular tie-rods (with length-to-diameter ratio less than 4) that connect the steam turbine rotors.
Their results revealed that the interaction between the horseshoe vortices near the upstream end wall and the wake shed from the cylinder varies with the inclination.
They also observed a jet-like stream in the cylinder wake that impinges on the downstream wall, and enhances the heat transfer. 
\citet{hu2015large} carried out large eddy simulations to study the flows over cable bridge pylon in the form of finite inclined square prism with height-to-width ratio of 18. 
In the forward inclination case, a dipole wake consisting a pair of counter-rotating tip vortices is formed, and is related to the strong downwash in the near wake.
In the backward inclination case, the wake is dominated by base vortex pair, while the tip vortices are suppressed.
\citet{zhang2020laminar} and \citet{zhang2022laminar} studied the laminar wake dynamics of finite-aspect-ratio swept wings at high angles of attack.
For both forward- and backward-swept wings, the steady vortices formed near the end boundaries not only stabilize the flow, but also provide additional vortex lift that enhances the aerodynamic performance.
The strong end effects described in these researches significantly influence the wake dynamics of low-aspect-ratio inclined cylinders (and prims, wings), and invalidate the independence principle.

Relatively less attention has been paid to the end boundary effects on flows over long inclined cylinders.
Interests in such configurations arise in the flows over trailing antenna, yawed stay cables, deepwater risers, just to name a few. 
Among the few who have addressed this issue, \citet{ramberg1983JFM} showed that wake can exhibit various flow patterns that are dictated by the end conditions.
With low inclination angle, the wake divides into multiple shedding modes with different angles of vortex tubes shed at distinct frequencies.
As inclination angle increases, the unsteady part of the wake moves away from the upstream boundary, and the majority of the span features steady trailing vortices on alternating sides of the cylinder with opposite circulation. 
Such steady flow pattern has also been observed in the wakes of slender cylindrical bodies at large incidence \citep{allen1951characteristics,sarpkaya1966separated,thomson1971spacing,zhang2020laminar}.
The complex flow features described above are significantly different from those found in the wakes of inclined infinite cylinders.

The wake dynamics of inclined cylinders exhibits highly three-dimensional characteristics originated from various mechanisms such the end effects, the Reynolds number effects, as well as the inclination angle.
While a lot of fascinating flow physics has been uncovered in the early work of \citet{ramberg1983JFM}, the roles played by the different mechanisms in the formation the three-dimensional flows are not clear.
The evolution of the vortical structures along the span is not described in detailed in previous studies.
In addition, the implications of the three-dimensional flow structures on the forces on the structures are yet to be unveiled.
The answers to these questions provide renewed insights into the complex wake dynamics observed in the key work of \citet{ramberg1983JFM}, and lay the foundation for the design of long flexible engineering structures and related flow control devices.

To address the above problems, we carry out direct numerical simulations to study the flows over circular cylinders with inhomogeneous end boundary conditions.
We cover three representative Reynolds numbers,  $\Rey=100$, $200$ and $300$, corresponding to the different flow regimes of two-dimensional laminar wake, mode A* wake, and mode B wake, respectively.
Relatively large length-to-diameter ratio is considered to allow the end-condition-induced three dimensionality to fully develop along the cylinder span.
These settings allow for a comprehensive understanding of the interplay between the extrinsic end boundary effects and intrinsic three dimensionality. 	
In what follows, we present the computational setup in \S \ref{sec:setup}. 
The numerical results are discussed in \S \ref{sec:results}, where we analyze the wake vortical structures and the corresponding hydrodynamic forces at different Reynolds numbers in detail.
We conclude this study by summarizing our findings in \S \ref{sec:conclusions}.

\section{Computational setup}
\label{sec:setup}

The flows over the circular cylinders are studied numerically by solving the three-dimensional Navier-Stokes equations:
\begin{equation}
\begin{split}
\boldsymbol{\nabla}\cdot \boldsymbol{u} & = 0,\\
\frac{\partial \boldsymbol{u}}{\partial t} + \boldsymbol{u}\cdot\boldsymbol{\nabla}\boldsymbol{u} & = -\boldsymbol{\nabla}p + \frac{1}{Re}\boldsymbol{\nabla}^2\boldsymbol{u},
\end{split}
\end{equation}
in which $\boldsymbol{u}=(u_x,u_y,u_z)$ is the velocity vector, and $p$ is the pressure.
The Reynolds number is defined as $\Rey\equiv U_{\infty}D/\nu$, where $U_{\infty}$ is the freestream velocity, $D$ is the diameter of the cylinder, and $\nu$ is the kinematic viscosity of the fluid. 
This study considers three representative Reynolds numbers, i.e., $\Rey=100$, 200 and 300, which lie in the regimes of two-dimensional flow, the mode A* wake, and the mode B wake.

\begin{figure}
\centering
\includegraphics[width=0.8\textwidth]{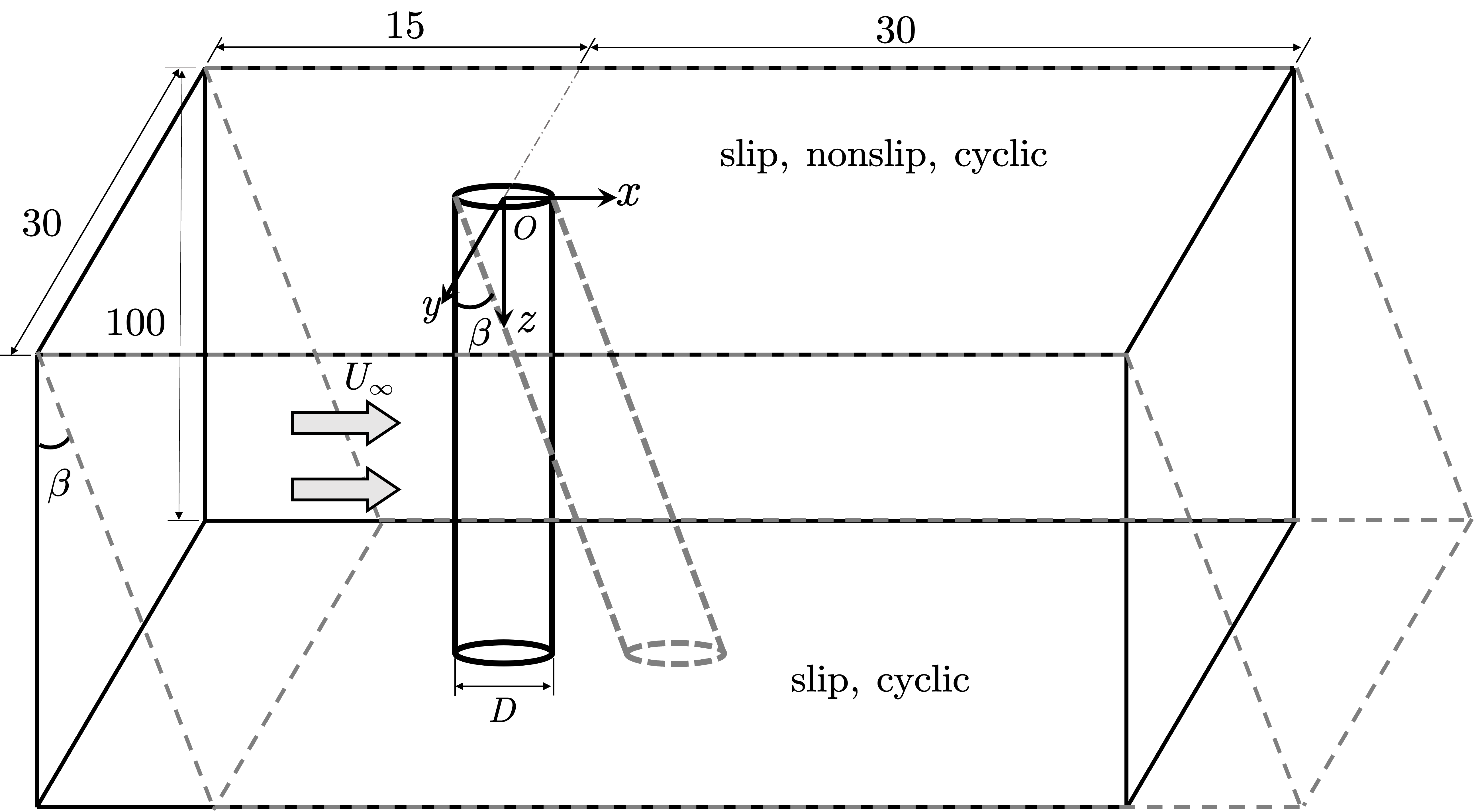}
\caption{Schematic of the computational setup. The black solid lines and gray dashed lines represent the setups for flows over straight ($\beta=0^{\circ}$) and inclined ($\beta\neq 0^{\circ}$) cylinders, respectively. The sketch is not to scale.}
\label{fig:setup}
\end{figure}

A schematic of the computational setup is shown in figure \ref{fig:setup}. 
The center of the circular cylinder is located at $15D$ from the inlet, and $30D$ from the outlet.
The width of the computational domain is $30D$, which results in a blockage ratio of 3.3\%.
The length of the cylinder in the $z$ direction is fixed at $100D$. 
The circumference of the cylinder is discretized with 160 grids. 
The mesh is concentrated in the vicinity of the cylinder to better resolve the near wake. 
The height of the first-layer mesh on the cylinder wall is $0.02D$. 
The spanwise direction is discretized with 2000 grids, resulting in a uniform spanwise grid resolution of $\Delta z = 0.05D$.
The total number of the grid numbers amounts to $\sim 4.6\times 10^7$.
The time-step is fixed at $\Delta t= 0.01 D/U_{\infty}$, resulting in a maximum CFL number below 1.
To simulate the flows over inclined cylinders, the computational domain is sheared with the inclination angle $\beta$ in the $x$ direction as shown by the dashed frame in figure \ref{fig:setup}, while the direction of the freestream velocity is kept fixed.
Note that in such setup, the incoming flow ``sees'' the section of the inclined cylinder still as circular shape.
In what follows, all variables are reported in their dimensionless forms by normalizing spatial quantities with the cylinder diameter $D$, velocity with $U_{\infty}$, and time with $D/U_{\infty}$.

\begin{table}
\renewcommand*{\arraystretch}{1.2}
 \begin{center}
  \begin{tabular}{l@{\hskip 0.5cm}|@{\hskip 0.5cm}c@{\hskip 1cm}c@{\hskip 1cm}c}
    		& $\beta=0^{\circ}$  &   $\beta=15^{\circ}$  &   $\beta=30^{\circ}$   \\
    		\midrule
       $\Rey=100$   & \textit{slip}, nonslip & \textit{cyclic}, nonslip, slip & \textit{cyclic}, nonslip, slip  \\
       $\Rey=200$   & \textit{slip}, nonslip & \textit{cyclic}, slip & \textit{cyclic}, slip \\
				$\Rey=300$   & \textit{slip}, nonslip & \textit{cyclic}, slip & \textit{cyclic}, slip  \\
  \end{tabular}
  \caption{End boundary conditions for $z=0$ plane. Italicized cases are considered spanwise homogeneous.}
 \end{center}
 \label{table:BC}
\end{table}

The finite-volume-based incompressible solver \emph{pimpleFoam} of the open-source CFD toolbox OpenFOAM \citep{weller1998tensorial} is used to carry out the direct numerical simulations. 
Both space and time are discretized with second-order accurate schemes. 
Different boundary conditions, including slip ($u_z=0$), nonslip ($\boldsymbol{u}=0$) and cyclic ($\boldsymbol{u}|_{z=0}=\boldsymbol{u}|_{z=100}$) are imposed on the $z=0$ plane.
The other side of the spanwise boundary ($z=100$), which is positioned downstream in the inclined cases, is generally treated as slip, except for the cyclic condition used together with that at $z=0$.
A summary of the simulated cases with different boundary conditions are shown in table \ref{table:BC}.
These cases allow us to inspect the flow physics in both spanwise homogeneous and inhomogeneous settings.
The rest of the boundary conditions are specified as follows.
The inlet is prescribed with a uniform inflow velocity $\boldsymbol{u}=(U_{\infty}, 0, 0)$. 
The side walls of the computational domain are treated as slip boundaries. 
Nonslip boundary condition is applied to the cylinder surface. 
A reference pressure $p_{\infty}=0$ is set at the outlet domain, with a zero-gradient condition for velocity. 

\begin{table}
\renewcommand*{\arraystretch}{1.2}
 \begin{center}
\begin{tabular}{l@{\hskip 0.4cm}|@{\hskip 0.4cm}c@{\hskip 0.4cm}c@{\hskip 0.4cm}c}
           								& $\Rey=100$ & $\Rey=200$ & $\Rey=300$ \\ 
\midrule
\citet{jiang2016three} 			& ---  	& ---   	& 1.30                 \\
\citet{canuto2015two} 			& 1.34 	& ---   	& ---                  \\
\citet{qu2013quantitative} 	& 1.32  	& 1.25  	& ---                  \\
Present    								& 1.35  	& 1.27 	& 1.31                 \\ 
Refined    								& 1.35  	& 1.26 	& 1.31               
\end{tabular}
  \caption{Comparison of the time-averaged drag coefficient $\overline{C_D}$ for flow over a two-dimensional circular cylinder at $\Rey=100$, 200 and 300. The refined case employs increased spatial grid numbers in all three spatial directions by a factor of 1.4.}
 \end{center}
 \label{table:validation}
\end{table}

\begin{figure}
\centering
\includegraphics[scale=0.4]{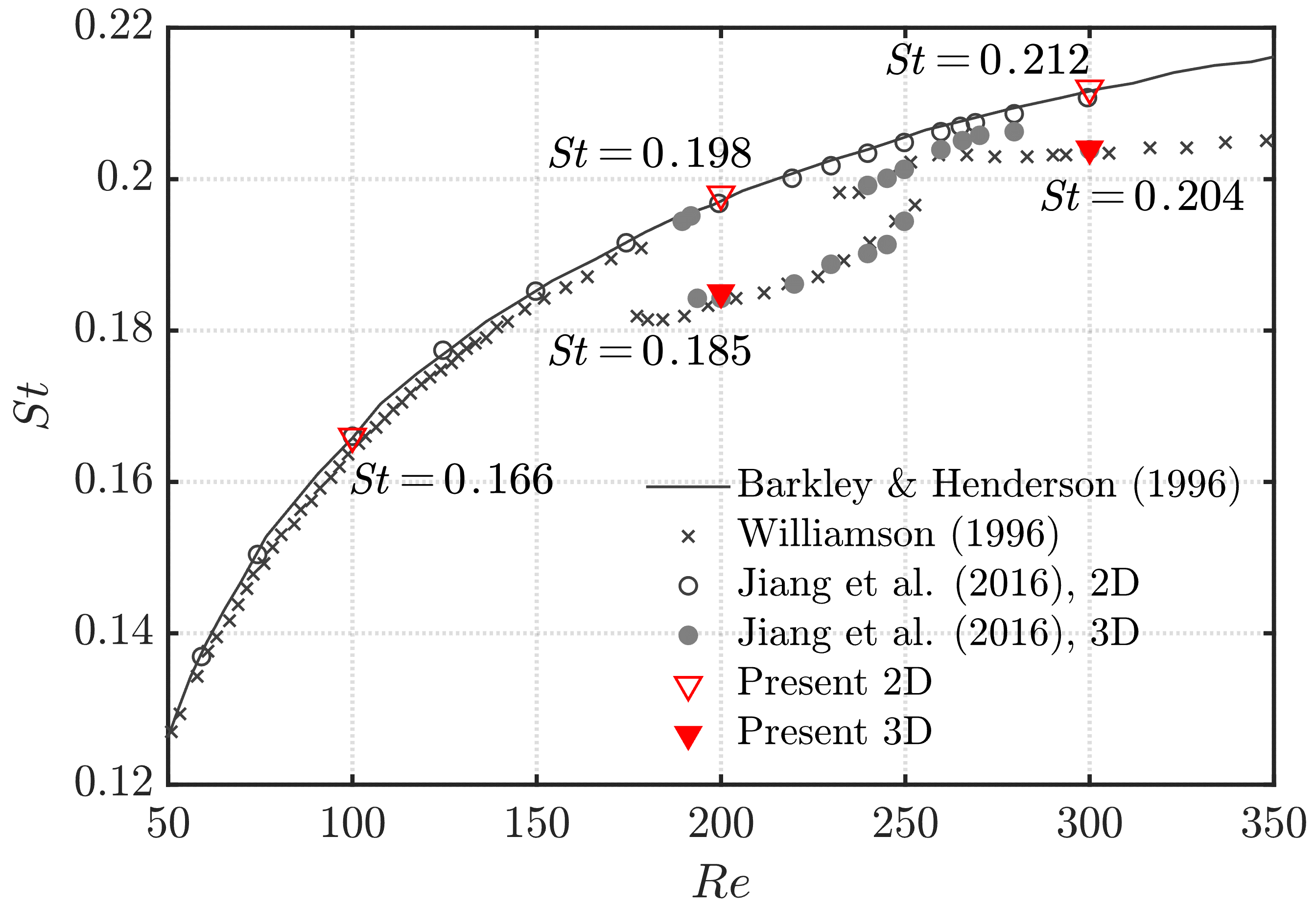}
\caption{Comparison of Strouhal number of flow over a circular cylinder at different Reynolds numbers. The present 2D simulations are carried out using one cell in the spanwise direction, thus suppressing the development of the intrinsic three dimensionality. The present 3D simulations refer to the spanwise homogenous cases with developed intrinsic three dimensionality.}
\label{fig:StCompare}
\end{figure}

To validate the above computational setup, we present a comparison of the time-averaged drag coefficient $\overline{C_D}$ for flows over spanwise-homogeneous circular cylinders at different Reynolds numbers in table \ref{table:validation}.
The drag and lift coefficients are defined as $C_D=F_D/(\rho U_{\infty}^2 DH/2)$ and $C_L=F_L/(\rho U_{\infty}^2 DH/2)$, where $F_D$ and $F_L$ are the drag and lift forces on the cylinder, and $H$ is the length of the cylinder.
It is observed that fair agreement between our numerical results and those from previous studies is achieved, demonstrating the accuracy of the current numerical method.
The results have also reached mesh independency, as a 1.4-fold refinement in grid resolution results in negligible changes in the drag coefficients.
We further compare the the Strouhal number (defined as $St=fD/U_{\infty}$, where $f$ is the vortex shedding frequency) of with \citet{jiang2016three}, \citet{barkley1996three} and \citet{williamson1996ARFM} in figure \ref{fig:StCompare} for the spanwise homogeneous cases.
The different Strouhal numbers in 2D and 3D simulations at $\Rey=200$ and 300 are accurately predicted. 

\section{Results}
\label{sec:results}
To reveal the different roles of the three-dimensional mechanisms play in the wake dynamics of inclined cylinders, we first present the numerical results on the extrinsic end boundary effects on the cylinder wakes at $\Rey=100$, where the intrinsic three dimensionality is absent.
The second part of this section discusses the mode A* ($\Rey=200$) and mode B ($\Rey=300$) wakes of both straight and inclined cylinders without the end boundary effects.
At last, the flows subjected to both the extrinsic and intrinsic three dimensionality are presented.

\subsection{End boundary effects at $\Rey=100$}
\label{sec:Re=100}

\subsubsection{Wake vortical structures}
\begin{figure}
\centering
\includegraphics[width=0.99\textwidth]{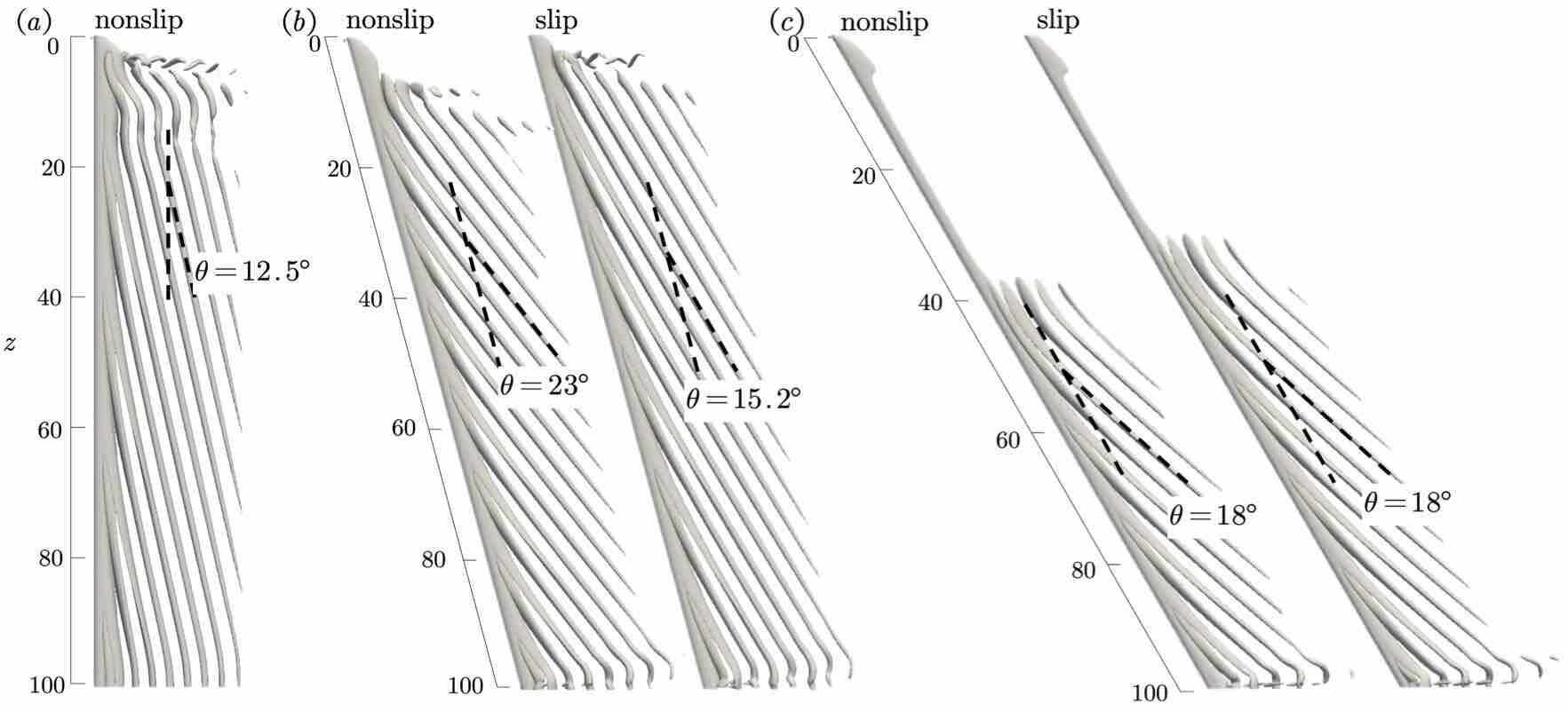}
\caption{Flows over circular cylinders at $\Rey=100$. ($a$) $\beta=0^{\circ}$, ($b$) $\beta=15^{\circ}$ and ($c$) $\beta=30^{\circ}$.
In $(b)$ and $(c)$, the results from both the nonslip and symmetry boundaries are shown. The vortical structures are visualized using isosurfaces of $QD^2/U_{\infty}^2=0.1$.}
\label{fig:Re100Vortical}
\end{figure}

The flows at $\Rey=100$ is generally laminar and offer a ``clean" condition to study the extrinsic end boundary effects.
The wake vortical structures visualized by isosurfaces of $QD^2/U_{\infty}^2=0.1$ ($Q$ is the second invariant of velocity gradient tensor) for flows at $\Rey=100$ are shown in figure \ref{fig:Re100Vortical}.
Under the nonslip end condition, the wake for the $\beta=0^{\circ}$ case exhibits typical cellular structures that have been described in previous studies \citep{behara2010flow,zhang2020formation,mittal2021cellular}.
The first cell near the nonslip wall occupies $0<z\lesssim 15$, and is featured by highly three-dimensional flow structures, including the braid-like vortices that mainly consists of streamwise and crossflow vorticity ($\omega_x$ and $\omega_y$), and the curved spanwise vortex tubes. 
It has been shown that the braid-like vortices connects the spanwise vortical structures to form vortex loops \citep{taneda1952studies,zhang2020formation,mittal2021cellular}. 
The second cell, which dominates most of the cylinder span, features oblique vortices with $\theta=12.5^{\circ}$, where $\theta$ is termed the oblique angle denoting the angle between the vortex tube and cylinder axis.
The vortex shedding frequency in the former cell is smaller than that of the second cell, thus creating vortex dislocation at $z\approx 15$.
Near the $z=100$ plane, the inclined vortex tubes are bent to be parallel to $z$ axis to respect the symmetry boundary condition.

For the inclined cases, we assess the effects of the nonslip and slip boundaries, both of which impose inhomogeneous end conditions to the K\'arm\'an vortex shedding.
At $\beta=15^{\circ}$, the wakes with both boundary conditions are dominated by vortex shedding obliqued at an angle to the already inclined cylinder axis, as observed in figure \ref{fig:Re100Vortical}($b$).
For the nonslip case, the vortex tubes are aligned at an angle of $\theta=23^{\circ}$ to the cylinder axis, while for the slip case, the oblique angle is $\theta=15.2^{\circ}$.
This comparison suggests the high sensitivity of the oblique angle on end condition at low inclination angles.
Similar observation has also been reported in \citep{behara2010flow} for the straight cylinders, in which the shedding angle is shown to be dependent on the boundary layer thickness of the nonslip wall, and the aspect ratio of the cylinder. 
In fact, \citet{mittal2014steady} and \citet{helder2022wing} have shown that for straight cylinder, there exists a vast number of unstable eigenmodes for the wake, each corresponding to a certain angle of oblique vortices. 
Aside from the difference in the oblique angle between the two cases, the nonslip wall induces a larger region of steady flow near the boundary than the slip wall. 
As the inclination angle increases to $\beta=30^{\circ}$, the steady region expand drastically to $z\approx 35$ -- 40.
An oblique shedding angle of $\theta=18^{\circ}$ is observed for both boundaries, suggesting its weak dependence on the end conditions at large inclination angle. 

\begin{figure}
\centering
\includegraphics[width=0.99\textwidth]{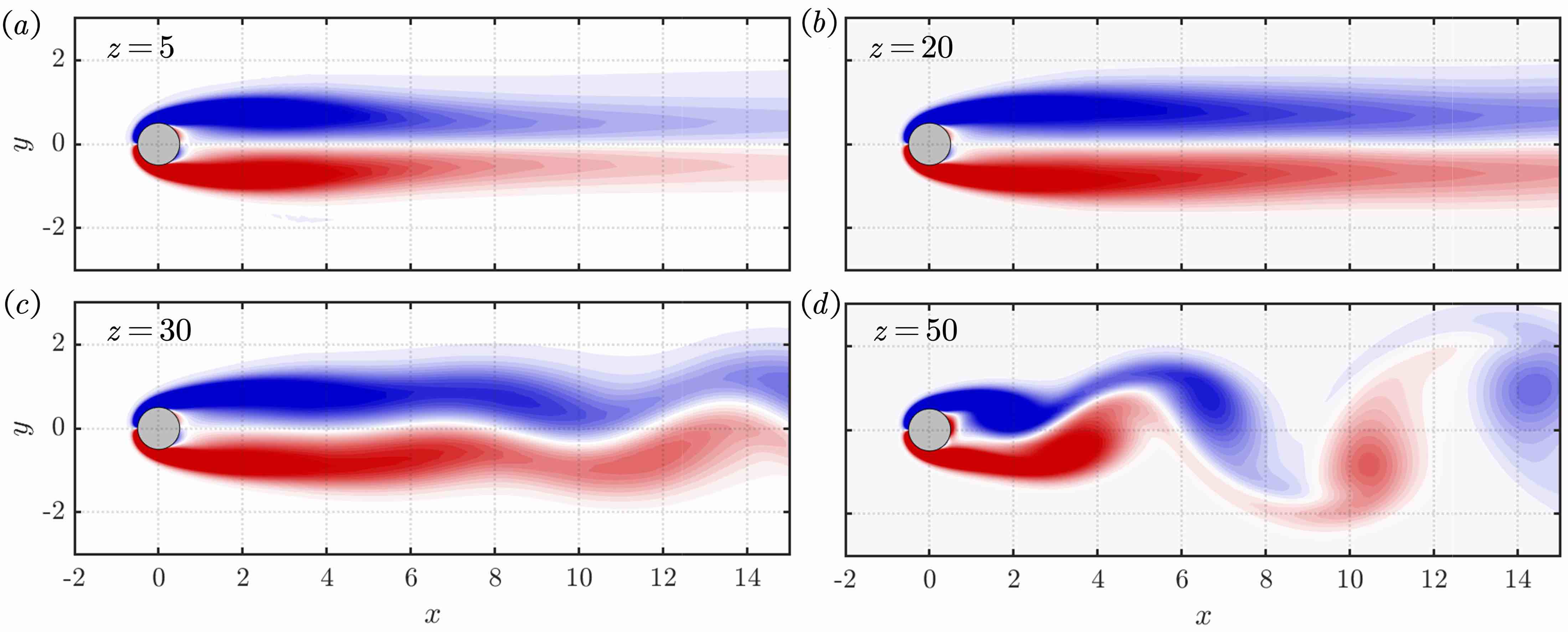}
\caption{Instantaneous $\omega_z$ fields for the case $(Re,\beta)=(100,30^{\circ})$ with slip end condition. Slices are taken at ($a$) $z=5$, ($b$) $z=20$, ($c$) $z=30$ and $(d)$ $z=50$. The $x$ coordinates of these slices are shifted so that the cylinder center resides at $(x,y)=(0,0)$. Blue and red indicate negative and positive $\omega_z$ in the range of $[-1,1]$. }
\label{fig:VortOnXY}
\end{figure}

\begin{figure}
\centering
\includegraphics[width=0.99\textwidth]{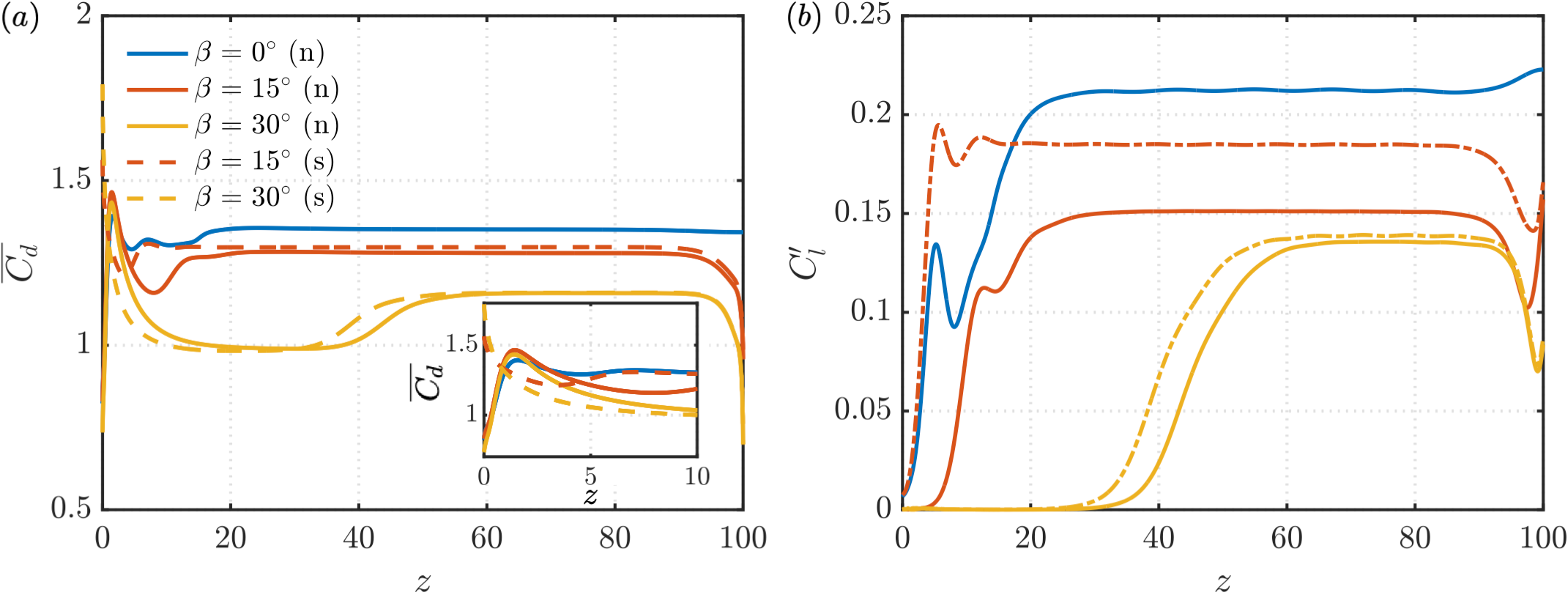}
\caption{Sectional force coefficients for cases with $\Rey=100$. ($a$) time-averaged drag coefficient and ($b$) rms lift coefficient. `n' and `s' represent nonslip and slip boundary conditions.}
\label{fig:sectionalCdClStatistics}
\end{figure}

Let us take a closer look at the transition from the locally steady flow to the unsteady vortex shedding in the case of $\beta=30^{\circ}$. 
As shown in figure \ref{fig:VortOnXY}, at $z=5$, the sectional flow (visualized by the $z$-component of vorticity vector $\boldsymbol{\omega}=\nabla\times\boldsymbol{u}$) is featured by a pair of counter-rotating vortices in the near wake.
The vorticity advects to further wake at $z=20$.
At $z=30$, the flow in the far wake starts to oscillate, and at $z=50$, the flow becomes fully unsteady featuring periodic shedding in the near wake.
The progressive development of the flow along the spanwise direction as described above is similar to the temporal evolution of the two-dimensional wake of a symmetric bluff body started impulsively from rest \citep{braza1986numerical,tamaddon1994unsteady,laroussi2014triggering}.
This is also reflected in the sectional force coefficients depicted in figure \ref{fig:sectionalCdClStatistics}.
For the case of $\beta=30^{\circ}$, the steady flow region is associated with lower drag and zero lift, while the unsteady shedding region corresponds to higher drag and lift forces.
The transition between the two states is similar to the destabilization of the flow from unstable steady state to the saturated unsteady flow \citep{noack2003hierarchy}.
Similar impulsive flow analogy has also been reported for the wakes of slender cylindrical bodies with large incidence \citep{allen1951characteristics,sarpkaya1966separated,thomson1971spacing,zhang2020laminar}, although in their cases the flow is steady.

We note that the flow characteristics in the direct vicinity of the boundary walls can be significantly different depending on the end conditions.
This is reflected by the distinct behaviors of the sectional drag coefficients as shown in figure \ref{fig:sectionalCdClStatistics}.
With nonslip wall, the sectional drag is low on the $z=0$ plane and then increases up to $z\approx 1$.
In the case of slip wall condition, the $C_d$ is high near the end, and decrease monotonically in the spanwise direction.
Despite the difference near the wall, the flows far away from the end boundaries feature oblique vortex shedding.
In view of this, we only adopt the slip boundary condition in studying the wake dynamics of inclined cylinders at higher Reynolds numbers, as will be discussed in \S \ref{sec:combined}.

\begin{figure}
\centering
\includegraphics[width=0.99\textwidth]{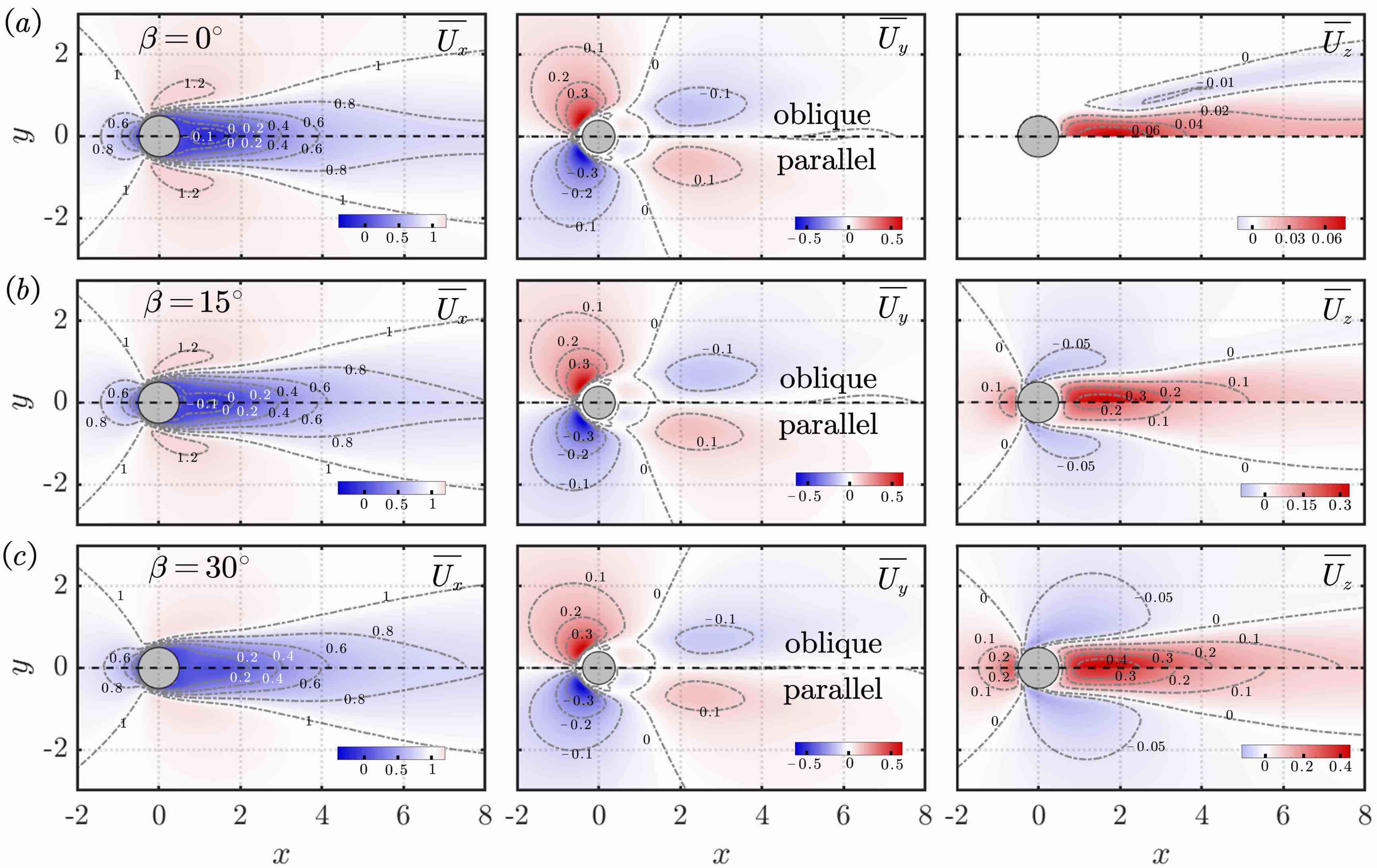}
\caption{Time averaged $z$-component vorticity (left), $x$-component vorticity (middle), and $z$-component velocity (right) fields for parallel and oblique (with nonslip boundary condition) vortex shedding at $\Rey=100$. $(a)$ $\beta=0^{\circ}$, $(b)$ $\beta=15^{\circ}$ and $(c)$ $\beta=30^{\circ}$.}
\label{fig:UMeanFields}
\end{figure}

We further assess the similarity of the flows in parallel and oblique vortex shedding from the time-averaged fields shown in figure \ref{fig:UMeanFields}.
From the $\overline{\omega_z}$ fields viewed on the $x$-$y$ plane, the flows are characterized by a pair of counter-rotating vortex cores, and not much difference is discerned in the mean flow between parallel and oblique shedding. 
This observation is in accordance with \citet{mittal2014steady}, in which the authors reported that the linear stability modes of parallel and oblique shedding of straight cylinders are similar.
However, the oblique shedding is associated with noticeably stronger streamwise vorticity $\omega_x$, and is also related to intensified spanwise flow, as depicted in the $\overline{u_z}$ fields.
The steep gradient of $u_z$ in the $y$ direction contributes to stronger streamwise vorticity component, and produces a slanted vorticity vector (with respect to the $z$-component) leading to oblique shedding.

\subsubsection{Spatio-temporal periodicity of the oblique vortex shedding}

\begin{figure}
\centering
\includegraphics[width = 0.7\textwidth]{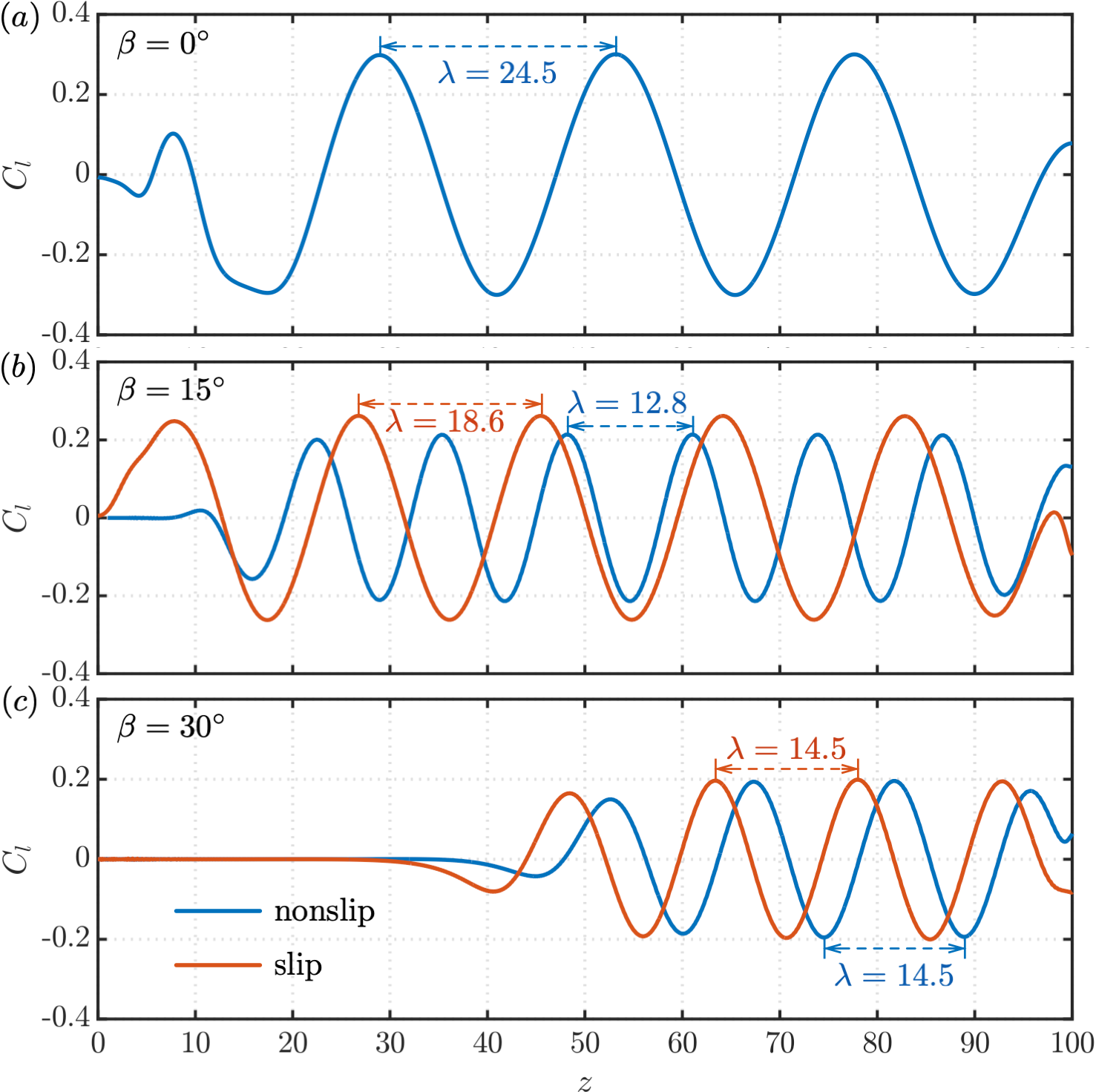}
\caption{Instantaneous sectional lift coefficients along the cylinder span. $\lambda$ denotes the spanwise wavelength of the oblique shedding.}
\label{fig:SpanwiseWavelength}
\end{figure}

The oblique vortex shedding is associated with periodic flow-induced forces in both space and time.
We present the instantaneous sectional lift coefficients along the span in figure \ref{fig:SpanwiseWavelength}.
For the spanwise regions occupied by the oblique vortex shedding, the sectional lift exhibits clear spatial periodicity that is characterized by the wavelength $\lambda$.
In general, larger oblique angle is associated with smaller wavelength, and vice versa.
Within each wavelength, the positive and negative lift coefficients cancel out, resulting in much weaker fluctuations in the integrated forces on the long cylinder.
The spatial wavelengths are presented in table \ref{table:CosineRule} as a summary.

\begin{table}
\centering
\renewcommand*{\arraystretch}{1.2}
\begin{tabular}{l@{\hskip 0.5cm}|@{\hskip 0.5cm}c@{\hskip 1cm}c@{\hskip 1cm}c@{\hskip 1cm}c@{\hskip 1cm}c@{\hskip 1cm}}
  									& b. c. 			 & $\theta$  		& $\lambda$     & $St$  & $St_{\rm{p},0}\cdot\cos(\beta+\theta)$ \\
  \midrule
\multirow{2}{*}{$\beta=0^{\circ}$}  & \textit{slip}    	 & $0^{\circ}$ 		&  $\infty$     & 0.167 & 0.167 \\
									& nonslip    		 & $12.5^{\circ}$ 	&  24.5     	& 0.163 & 0.163\\		
\midrule
\multirow{3}{*}{$\beta=15^{\circ}$} & \textit{cyclic}    & $0^{\circ}$    	&  $\infty$     & 0.160 & 0.161 \\
									& nonslip    		 & $23^{\circ}$   	&  12.8     	& 0.132 & 0.132 \\
									& slip    	 		 & $15.2^{\circ}$ 	&  18.6     	& 0.145 & 0.144 \\
\midrule
\multirow{3}{*}{$\beta=30^{\circ}$} & \textit{cyclic}    & $0^{\circ}$    	&   $\infty$    & 0.143 & 0.145 \\
									& nonslip    		 & $18^{\circ}$   	&  14.5     	& 0.110 & 0.112 \\
									& slip    	 		 & $18^{\circ}$   	&  14.5     	& 0.110 & 0.112
\end{tabular}
		\caption{Spatio-temporal periodicity of oblique shedding at $\Rey=100$. $\lambda$ is the spatial wavelength, and $St$ is the shedding frequency measured from DNS. $St_{p,0}=0.167$ is the Strouhal number for parallel shedding at $(Re,\beta)=(100,0^{\circ})$.}

	\label{table:CosineRule}
\end{table}

Temporally, the oblique shedding is characterized by the Strouhal number $St=fD/U_{\infty}$, as also presented in table \ref{table:CosineRule}.
In both cases of parallel shedding of inclined cylinders \citep{van1968experiments,lucor2003effects}, and oblique shedding of straight cylinders \citep{williamson1989oblique,mittal2021cellular}, the shedding frequency can be predicted by $St=St_{p,0}\cdot\cos\gamma$, where $St_{p,0}$ is the Strouhal number for parallel shedding of straight cylinders, and $\gamma$ represents either the inclination angle in the former case, or oblique angle in the latter case. 
However, it remains elusive how the cosine law can be applied to the case of oblique shedding in the wake of inclined cylinders.
We show in table \ref{table:CosineRule} that the using the combined angle $\beta+\theta$ with $St_{p,0}$, the predicted frequencies by cosine law are in good agreement with the Strouhal numbers measured from DNS.
This finding shows that the independence principle can be generalized to the oblique shedding in the inclined cylinders, and will prove useful in interpreting results at higher Reynolds numbers in \S \ref{sec:combined}.

\subsection{The intrinsic three dimensionality}
\label{sec:intrinsic}
\begin{figure}
		\centering
    \includegraphics[width=1\textwidth]{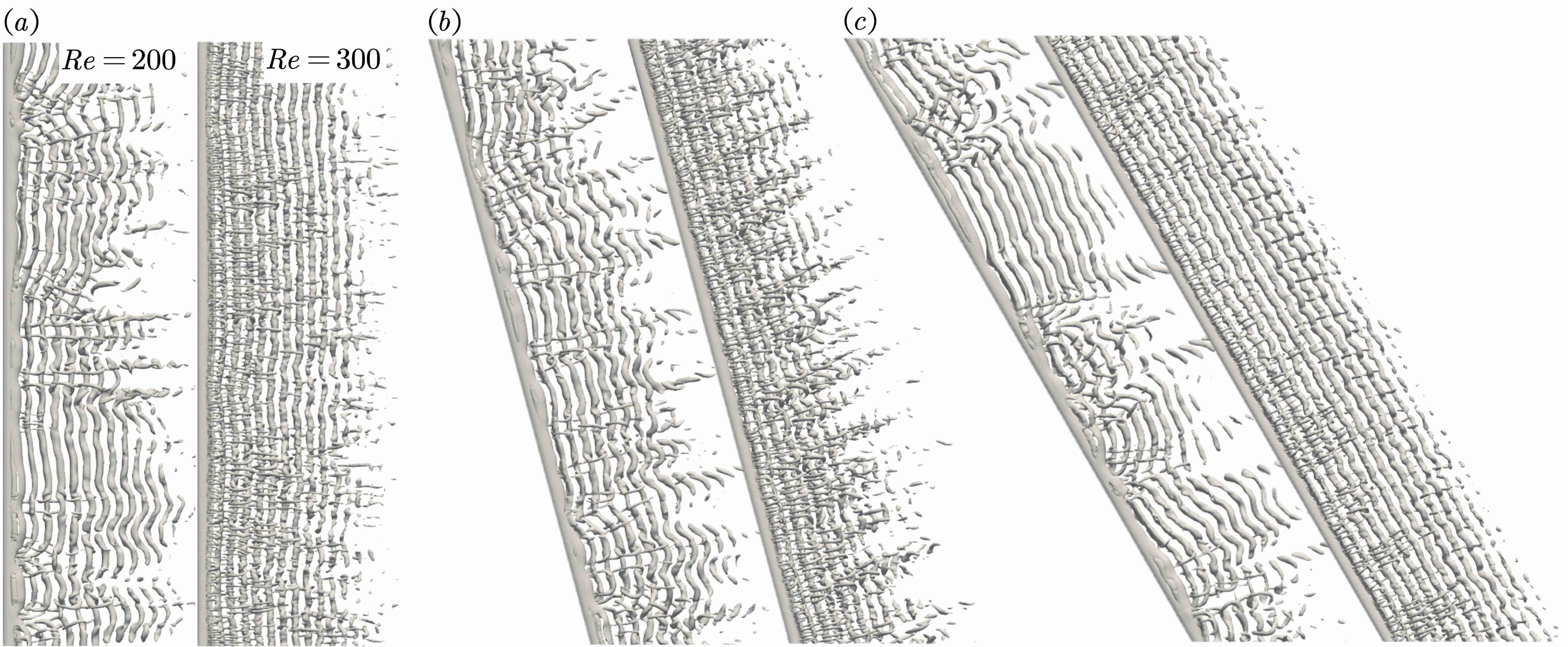}
    \caption{Flows over circular cylinders with homogeneous end condition at $\Rey=200$ (left) and 300 (right). ($a$) $\beta=0^{\circ}$, ($b$) $\beta=15^{\circ}$ and ($c$) $\beta=30^{\circ}$. Shown are isosurfaces of $QD^2/U_{\infty}=0.1$.}
    \label{fig:intrinsicQ}
\end{figure}

\begin{figure}
\centering
\includegraphics[width=1\textwidth]{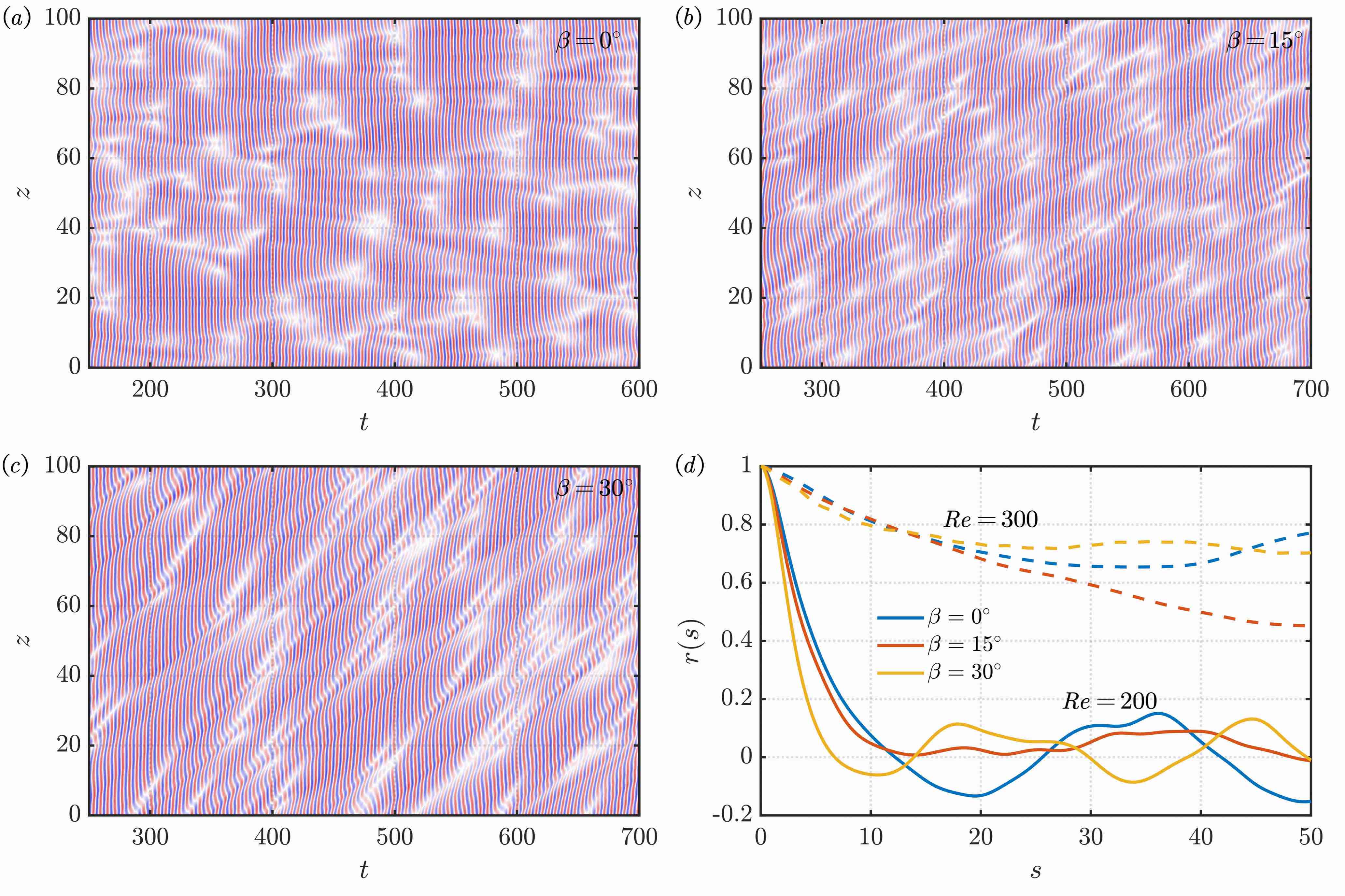}
\caption{Spatio-temporal distribution of sectional lift coefficients $C_l$ at $\Rey=200$ for ($a$) $\beta=0^{\circ}$, ($b$) $\beta=15^{\circ}$ , ($c$) $\beta=30^{\circ}$. Blue and red colors represent negative and positive $C_l$ in the range of $[-1, 1]$. ($d$) shows the averaged spanwise correlation coefficients of lift coefficients.}
\label{fig:correlation}
\end{figure}

At $\Rey=200$ and 300, secondary instabilities develop in the cylinder wake, giving rise to the intrinsically three-dimensional flows.
The wake vortical structures of spanwise homogeneous cylinders at these two Reynolds numbers are shown in figure \ref{fig:intrinsicQ} for different inclination angles.
For $\Rey=200$, the wakes of straight and inclined cylinders are featured by  vortex shedding with large-scale dislocations.
The occurrence of vortex dislocations induces locally bent vortex tubes, however, the K\'arm\'an vortex shedding are mostly parallel to the cylinder axis regardless of the inclination angle.
At $\Rey=300$, the flows are characterized by coherent K\'arm\'an vortex shedding interlaced with small-scale vortices that are mostly aligned perpendicular to the spanwise vortex tubes.
These features are in accordance with the descriptions of mode A* and mode B instabilities as described extensively in previous studies \citep{williamson1996JFM,williamson1996ARFM,braza2001successive,jiang2016three}.

To further assess the intrinsic three dimensionality in the cylinder wake, we present the spatio-temporal distributions of the sectional lift coefficients in figure \ref{fig:correlation}($a-c$) for $\Rey=200$ with different inclination angles.
For the straight cylinder, the spanwise coherence of vortex shedding is disrupted by the random occurrence of large-scale vortex dislocations, as represented by the white dots in figure \ref{fig:correlation}($a$).
The vortex dislocations are generally compact in space, but can developed over an extended range of time.
For $\beta=15^{\circ}$ and $30^{\circ}$, the vortex dislocations extend not only temporally, but also spatially, as evidenced by the stretched white threads in figure \ref{fig:correlation}($b,c$).
This observation indicates that the points of dislocations travel along the span, due to the spanwise flow in the inclined cases.

For spanwise-homogenous cases, the three dimensionality of the cylinder wake can be quantified using the spanwise correlation coefficient of the sectional lift coefficients, which is defined as 
\begin{equation}
r(z_i,z_j) = \frac{\overline{(C_{l,i}-\overline{C_{l,i}})(C_{l,j}-\overline{C_{l,j}})} }{\sqrt{\overline{(C_{l,i}-\overline{C_{l,i}})^2}}\sqrt{\overline{(C_{l,j}-\overline{C_{l,j}})^2}}}.
\label{equ:correlation}
\end{equation}
In this matrix, the elements in each of the diagonals parallel to the leading diagonal contains the correlation between two spanwise locations separated by the same distance $s$.
By averaging the elements in each diagonal, the information in the correlation coefficient matrix (\ref{equ:correlation}) is condensed to a single curve as a function of $s$.
The averaged spanwise correlation coefficients $r$ for the sectional lift coefficients are shown in figure \ref{fig:correlation}($d$). 
Due to the occurrence of large-scale vortex dislocations at $\Rey=200$, the spanwise vortex shedding loses coherence over small distance.
The spanwise correlation decreases more dramatically with $s$ for the inclined cases, due to the spanwise velocity component that prolongs the vortex dislocation process along the cylinder axis.
For $\Rey=300$, the spanwise correlation remains much higher than that at $\Rey=200$, suggesting that the intrinsically developed large-scale vortex dislocation is not prevalent.
For this reason, the spatio-temporal distributions of the sectional lift at $\Rey=300$ are not presented here for brevity.

\subsection{Combined effects of extrinsic and intrinsic three dimensionality}
\label{sec:combined}
\subsubsection{Boundary effects at $\Rey=200$}
\begin{figure}
\centering
\includegraphics[width=1\textwidth]{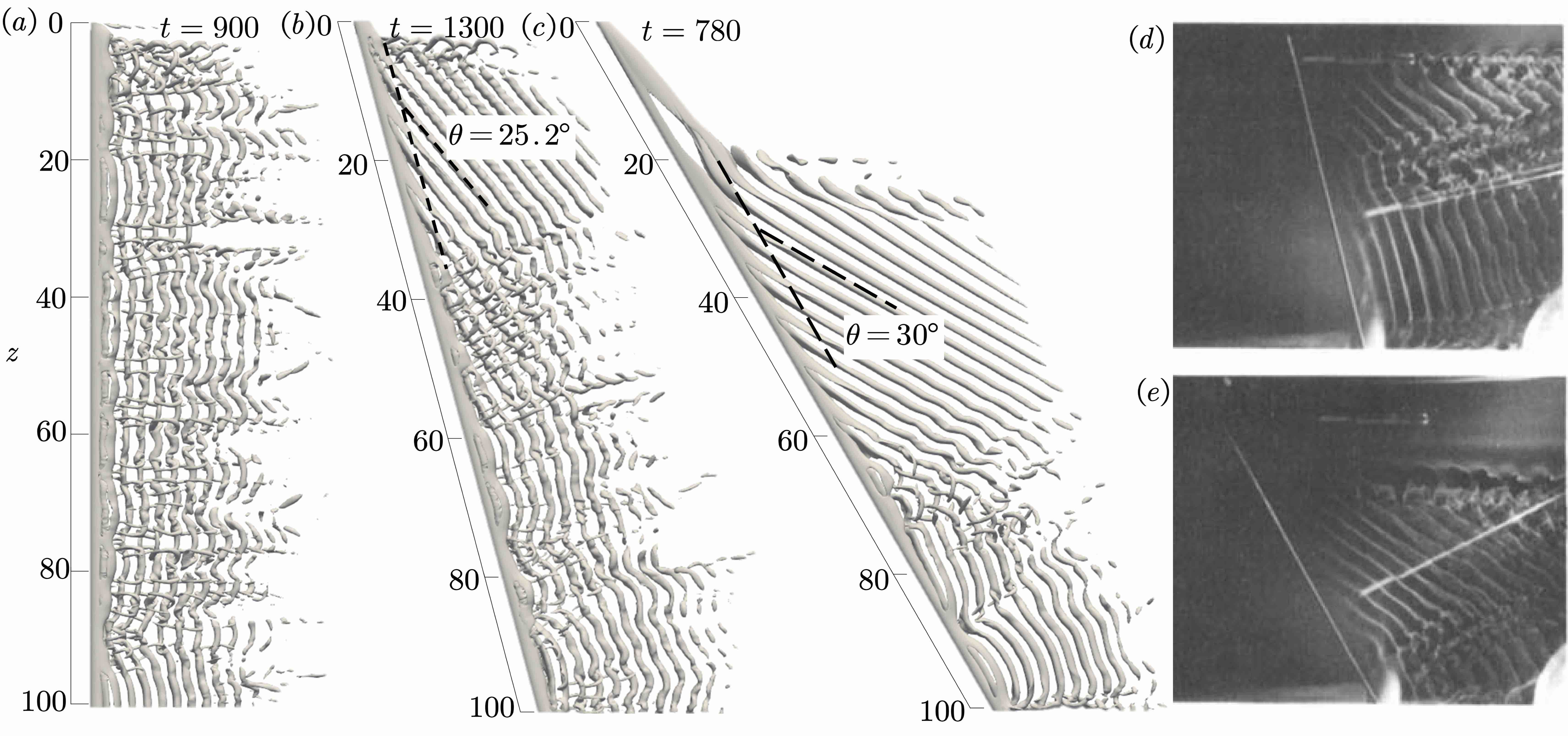}
\caption{Wake vortical structures for cylinders with inhomogeneous end conditions at $\Rey=200$. $(a)$ $\beta=0^{\circ}$, ($b$) $\beta=15^{\circ}$ and ($c$) $\beta=30^{\circ}$. Shown are isosurfaces of $QD^2/U_{\infty}=0.1$. $(d)$ and $(e)$ show the wake structures of inclined cylinders at $\Rey=160$  for $\beta=15^{\circ}$ and $30^{\circ}$, respectively. Reproduced from \citet{ramberg1983JFM} with permission from the Cambridge University Press.}
\label{fig:Re200EndEffects}
\end{figure}

\begin{figure}
\centering
\includegraphics[width=1\textwidth]{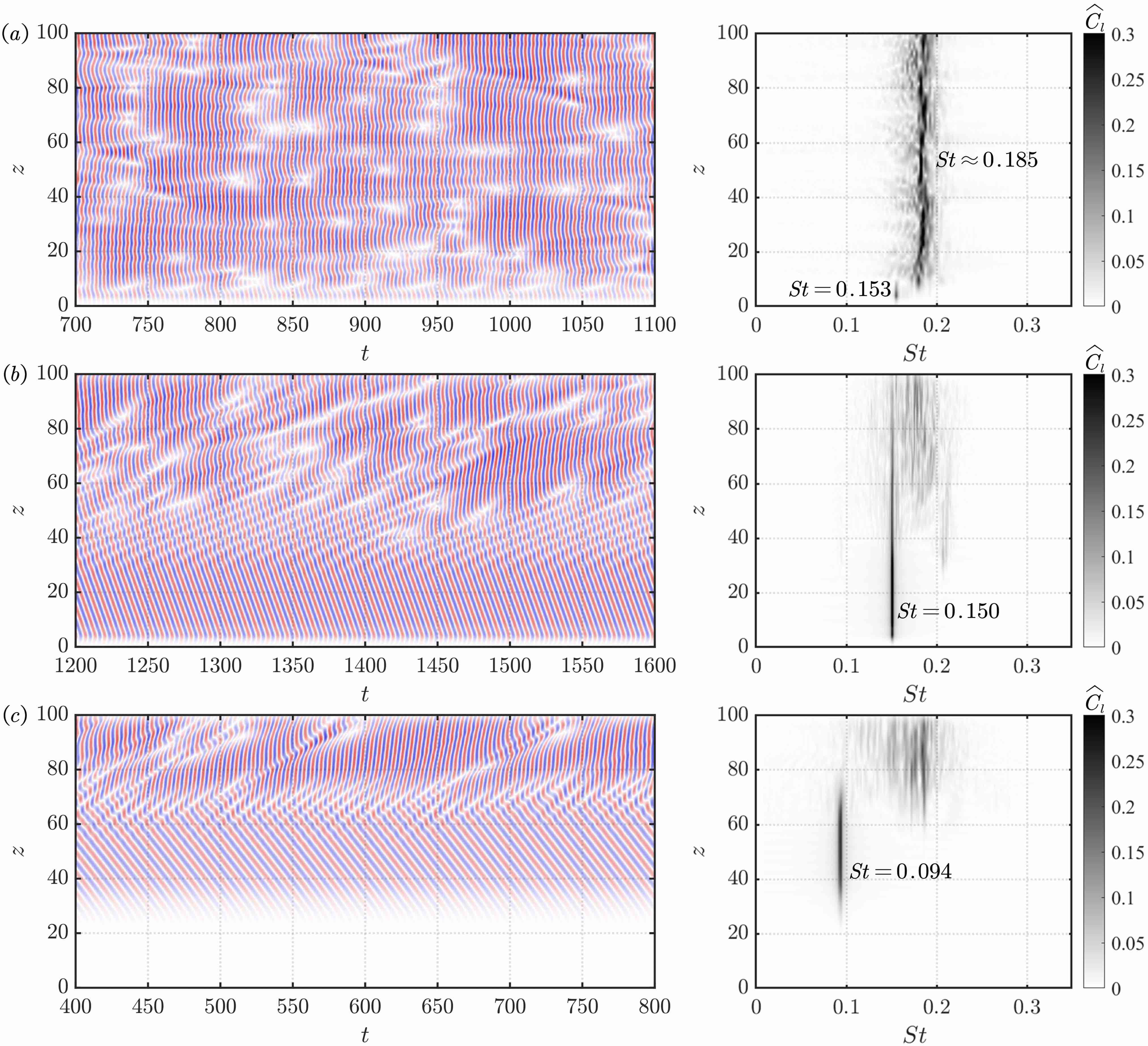}
\caption{Sectional lift coefficients (left) and their amplitude spectrum for cylinders with inhomogeneous end conditions at $\Rey=200$. $(a)$ $\beta=0^{\circ}$, ($b$) $\beta=15^{\circ}$ and ($c$) $\beta=30^{\circ}$.}
\label{fig:Re200Cl}
\end{figure}

In this section, we examine the flows over inclined cylinders at $\Rey=200$ and 300 to evaluate the roles of both extrinsic and intrinsic three dimensionality in shaping the wake dynamics.
The wake vortical structures for $\Rey=200$ with inhomogeneous end conditions are shown in figure \ref{fig:Re200EndEffects}.
The corresponding spatio-temporal distributions of sectional lift coefficients and their spectrum are shown in figure \ref{fig:Re200Cl}.
For the straight cylinder, the effects of nonslip boundary condition are localized near the nonslip boundary, where a layer of braid-like vortices with a low shedding frequency is observed.
The remaining part of the span features spanwise vortex tubes interlaced with large-scale vortex dislocation at a primary frequency of $St\approx0.185$, which is typical of the mode A* wake as described in \S \ref{sec:intrinsic}.
This differs from the scenario at $\Rey=100$ where the boundary effects are felt all over the span by inducing the oblique shedding.
This observation suggests that the occurrence of vortex dislocation in mode A* is able to preclude the propagation of the boundary effects to further span.
Similar observations were reported by \citet{prasad1997JFM}, in which the authors reported loss of flow control efficacy using end plates on the cylinder wake at $\Rey \approx 190-250$.

The wake dynamics for $\beta=15^{\circ}$ with inhomogeneous end boundary condition at $\Rey=200$ exhibit rich features along the span.
The immediate vicinity of the upstream boundary is featured by a region of steady flow, and is followed by the braid-like vortices up to $z\approx 5$.
For $5<z\lesssim 35$, organized oblique vortex shedding that is similar to the analogous case at $\Rey=100$ is observed.
These oblique vortices exhibit intact cores with slight waviness, but no vortex dislocation is developed.
The shedding frequency of these oblique vortices is $St=0.150$ as shown in figure \ref{fig:Re200Cl}($b$).
It is well acknowledged that the Strouhal numbers of the cylinder wake at $\Rey=200$ are different in 2D and 3D simulations \citep{williamson1988existence,jiang2016three}. 
The 2D simulation suppresses vortex dislocation, and results in a shedding frequency of $St_{2D}=0.196$, while the 3D simulation yields $St_{3D}=0.185$, as illustrated in figure \ref{fig:StCompare}. 
With the oblique angle $\theta=25.2^{\circ}$ at the inclination of $\beta=15^{\circ}$, the application of the cosine law using the two- and three-dimensional shedding frequencies gives $St_{2D}\cdot \cos(\theta+\beta)=0.150$ and $St_{3D}\cdot \cos(\theta+\beta)=0.141$, the former being in close agreement with the observed shedding frequency in figure \ref{fig:Re200Cl}($b$).
This suggests that the oblique shedding vortices observed here are related to the two-dimensional vortex shedding without the occurrence of vortex dislocation.

Towards the end of the oblique vortices, the vortex tubes break down with the formation of small-scale vortices that are mostly perpendicular to the oblique vortex cores.
This is reflected as more noticeable wavy patterns in the spatio-temporal lift distribution in $30\lesssim z \lesssim 60$, as shown in figure \ref{fig:Re200Cl}($b$).
This type of flow is reminiscent of the mode A wake with organized long-wavelength streamwise vortices, as described in \citep{zhao2013three,jiang2016three}.
Such mode-A-like wake develops from purely two-dimensional K\'arm\'an vortex shedding, and exists only for a finite period of time before it transitions to the mode A* wake with large-scale vortex dislocation.
In the wake of the inclined cylinder, such mode-A-like flow also occupies limited spanwise region.
From $z\approx 60$, the mode-A-like vortices destabilizes, and the conventional parallel vortex shedding with large-scale vortex dislocation dominates the wake, indicating the termination of end effects.
The development of the flow features in the spanwise direction, i.e., from the oblique vortex shedding with intact vortex cores, to the appearance of the mode-A-like flows, and finally to the mode A* flow along the cylinder span, presents another case of temporal-like flow evolution as previous described with regards to figure \ref{fig:VortOnXY}.

Further increasing the inclination angle to $\beta=30^{\circ}$, the immediate vicinity of the upstream boundary features steady flow up to $z\approx 20$. 
The organized oblique vortices are angled at $\theta=30^{\circ}$ to the cylinder axis, and shed at a frequency of 0.094, which is close to the prediction of the cosine law using the 2D frequency $St_{2D}\cdot\cos(\beta+\theta)=0.098$.
These oblique vortices break down at $z\approx 65$, and the flow transitions to parallel vortex shedding with large-scale vortex dislocation. 
In contrast to the scenario in $\beta=15^{\circ}$ where the mode-A-like flow is observed in between the organized oblique vortices and parallel vortices, the transition between these two flows appear more direct.
The general flow features revealed in the current direct numerical simulations are in good agreement with those observed in the early experiments by \citet{ramberg1983JFM}, as shown in figure \ref{fig:Re200EndEffects}($d,e$).
The combined effects of extrinsic inhomogeneous end boundary and intrinsic secondary instability produce highly complex flow physics involving oblique shedding, parallel shedding, vortex dislocation, cellular shedding, etc., all in a single wake.

\subsubsection{Boundary effects at $\Rey=300$}

\begin{figure}
\centering
\includegraphics[width=1\textwidth]{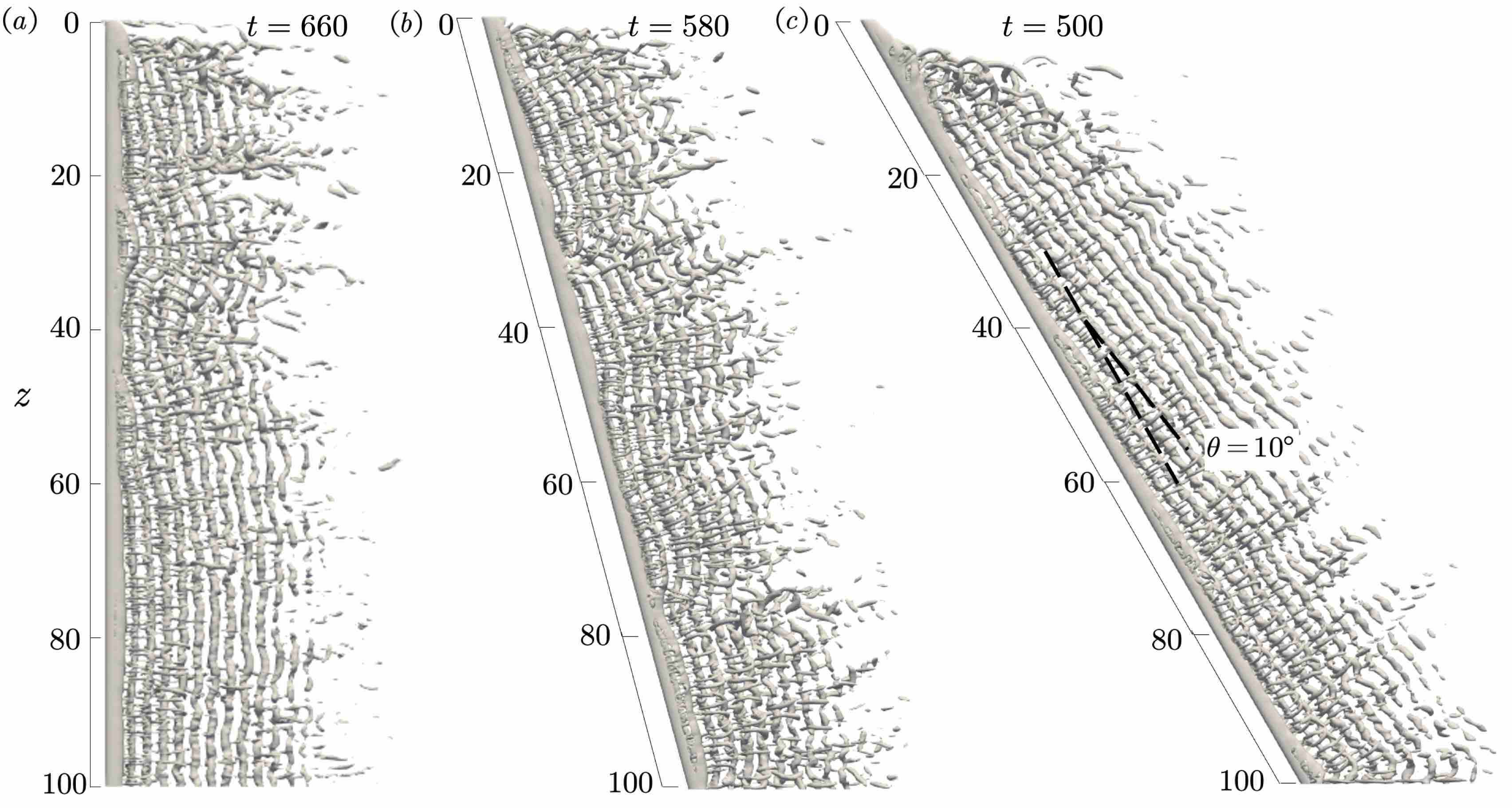}
\caption{Wake vortical structures for cylinders with inhomogeneous end conditions at $\Rey=300$. $(a)$ $\beta=0^{\circ}$, ($b$) $\beta=15^{\circ}$ and ($c$) $\beta=30^{\circ}$. Shown are isosurfaces of $QD^2/U_{\infty}=0.1$.}
\label{fig:Re300EndEffects}
\end{figure}

\begin{figure}
\centering
\includegraphics[width=0.99\textwidth]{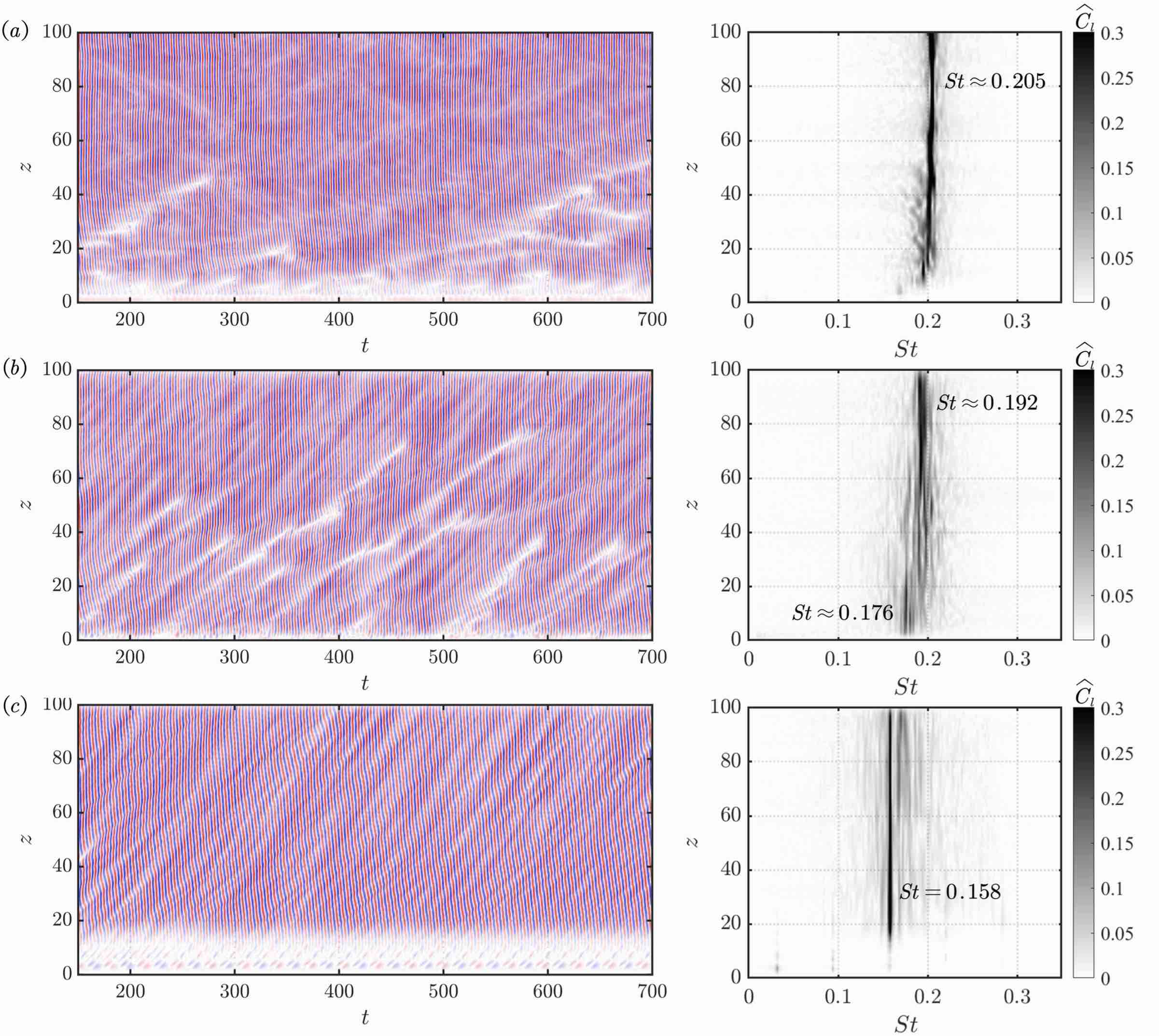}
\caption{Sectional lift coefficients for cylinders at $\Rey=300$. $(a)$ $\beta=0^{\circ}$ with nonslip end condition, ($b$) $\beta=15^{\circ}$ with slip end condition and ($c$) $\beta=30^{\circ}$ with slip end condition. Blue and red indicate positive and negative values ($C_l\in[-1,1]$), respectively.}
\label{fig:Re300Cl}
\end{figure}

The wake vortical structures of the inclined cylinders at $\Rey=300$ are presented in figure \ref{fig:Re300EndEffects}, with the corresponding spatio-temporal lift distributions and lift spectra shown in figure \ref{fig:Re300Cl}.
For $\beta=0^{\circ}$, the flow generally resembles the mode B wake as described in figure \ref{fig:intrinsicQ}, and the oblique vortex shedding that is observed at lower Reynolds numbers does not present a persistent feature at $\Rey=300$. 
Along the span, the coherence of the spanwise vortex tubes is disrupted by several instances of vortex dislocations (e.g., $z\approx 25$, 35, 45 at $t=660$).
Noticing that the large-scale vortex dislocation is a rare event in the spanwise homogeneous case as discussed in \S \ref{sec:intrinsic}, the dislocations observed in the current case should be considered as induced by the inhomogeneous boundary condition.
Indeed, from figure \ref{fig:Re300Cl}$(a)$, we observe that the occurrence of dislocations are mostly originated from the region near the nonslip boundary, and slowly moves inwards along the span.
For the simulated range of time, the propagation of the vortex dislocation is limited to $z\lesssim 50$, beyond which the vortex shedding is generally parallel to the cylinder axis.

Increasing the inclination to $\beta=15^{\circ}$, the vortex shedding near the inhomogeneous boundary is slightly obliqued to the cylinder axis. 
This is reflected by the smaller shedding frequency of $St\approx0.176$ at $0<z\lesssim20$ compared with the rest of the span.
Similar to the scenario at $\beta=0^{\circ}$, the disturbances generated near the boundary generate vortex dislocations that travel along the span.
With the spanwise flow behind the inclined cylinder, the dislocations propagate at a faster rate compared to the case of $\beta=0^{\circ}$, and in some cases they can penetrate a larger region in the spanwise direction. 
The vortex dislocations also result in broadband lift spectrum at $20\lesssim z \lesssim 70$.
Beyond $z\approx 80$, parallel vortex shedding associated with a primary frequency of $St\approx0.192$ becomes dominant.

At $\beta=30^{\circ}$, the direct vicinity of the upstream end boundary is occupied by steady flow, although this region is much shorter compared to the analogous cases at $\Rey=100$ and 200.
The rest of the span features oblique vortex shedding with $\theta\approx10^{\circ}$ and a primary frequency of $St=0.158$.
Towards the downstream end of the span, the lift spectra gradually becomes broadband.
However, no clear occurrence of vortex dislocation is observed in the flow.
We speculate that the lack of vortex dislocation at $\beta=30^{\circ}$ is due to the steady flow formed near $z=0$ plane.
Without the vibrant vortical activities near the end boundary triggering vortex dislocations as observed in the cases of $\beta=0^{\circ}$ and $15^{\circ}$, the boundary effects at $\beta=30^{\circ}$ manifest as oblique shedding similar to the scenario at $\Rey=100$.

\section{Conclusions}
\label{sec:conclusions}
We have carried out direct numerical simulations to study the wake dynamics of inclined long circular cylinders with spanwise-inhomogeneous boundary conditions at $\Rey=100$, 200 and 300.
The aim is to clarify the different roles of the end boundary conditions, the intrinsic secondary instabilities, and the inclination in the formation of the three-dimensional vortex shedding behind the cylinders.

At $\Rey=100$ where the intrinsic three dimensionality is absent, the effects of the inhomongeous end boundary are felt over the entire span, with the formation of oblique vortex shedding for both straight and inclined cylinders.
Compared with parallel shedding, the oblique shedding is associated with stronger spanwise velocity, which creates additional streamwise vorticity that results in oblique angle.
The oblique shedding features periodicity in both space and time.
Along the span, the positive and negative sectional lift coefficients cancel out within one wavelength, resulting in much smaller fluctuating lift over the entire cylinder.
Temporally, we found that the frequency of the oblique shedding is related to that of the parallel shedding of straight cylinder by a factor of $\cos(\beta+\theta)$, where $\beta$ is the inclination angle and $\theta$ is the oblique angle.
This observation indicates that the cosine law can be generalized to the oblique shedding in inclined cylinders.

At $\Rey=200$, the secondary instability develops in the cylinder wake as large-scale vortex dislocations, resulting in intrisically three-dimensional flow. 
For the straight cylinder with nonslip end condition, the occurrence of vortex dislocations precludes the propagation of the boundary effects towards further span.
For inclined cylinders, the inhomomgenous end boundary induces oblique shedding within a finite region along the span.
These oblique vortices are related to the intrinsically two-dimensional flows at $\Rey=200$ without large-scale vortex dislocations.
This is evidenced by the observation that the Strouhal number of the oblique shedding is in close agreement with that predicted by the cosine law, if the two-dimensional shedding frequency (with secondary instabilities suppressed) is used.
Further along the span, the intact oblique vortex cores destabilize with the formation of small-scale vortices, and the flow transitions to the typical mode A* wake.

At $\Rey=300$, the spanwise-homogeneous cylinder features the mode B wake with stable K\'arm\'an vortex shedding interlaced with fine-scale streamwise vortices.
With inhomogeneous end boundary conditions, the oblique shedding that is prevalent at lower Reynolds numbers does not present a persistent feature for cases with small inclination angle. 
The three-dimensional unsteady flow near the end boundary creates disturbances that propagate along the span, and later develops into vortex dislocation.
For high inclination angle, the oblique vortex shedding is again observed throughout the cylinder span, and is not disrupted by vortex dislocations of either intrinsic or extrinsic cause.

The above analysis have highlighted the complex interactions between the effects of extrinsic and intrinsic three dimensionality in forming the cylinder wake.
For flows without intrinsic instabilities, the extrinsic end effects propagate along the cylinder span unchallenged, resulting in oblique vortex shedding for cylinders even hundreds diameters in length.
The propagation of the end boundary effects can be terminated by the occurrence of large-scale vortex dislocations that develop intrinsically in the flow, giving rise to end-effects-free flow.
At higher Reynolds numbers, while the intrinsic three dimensionality in the form of mode B wake rarely creates vortex dislocations, the highly three-dimensional flow near the inhomogeneous boundary can be a source of perturbations that later develop into vortex dislocations.
The present study provides new insights into the three-dimensional wake dynamics of inclined cylinders.
The knowledge obtained here is fundamental to the understanding of higher-Reynolds-number flow and fluid-structure interactions of flexible bodies.
It also lays the foundation for the designs of large length-to-diameter-ratio engineering structures and related flow control techniques.

\section*{Appendix I: Cosine law for flow over inclined two-dimensional cylinders}
\label{sec:cosineLaw}

\begin{figure}
\centering
\includegraphics[width=0.6\textwidth]{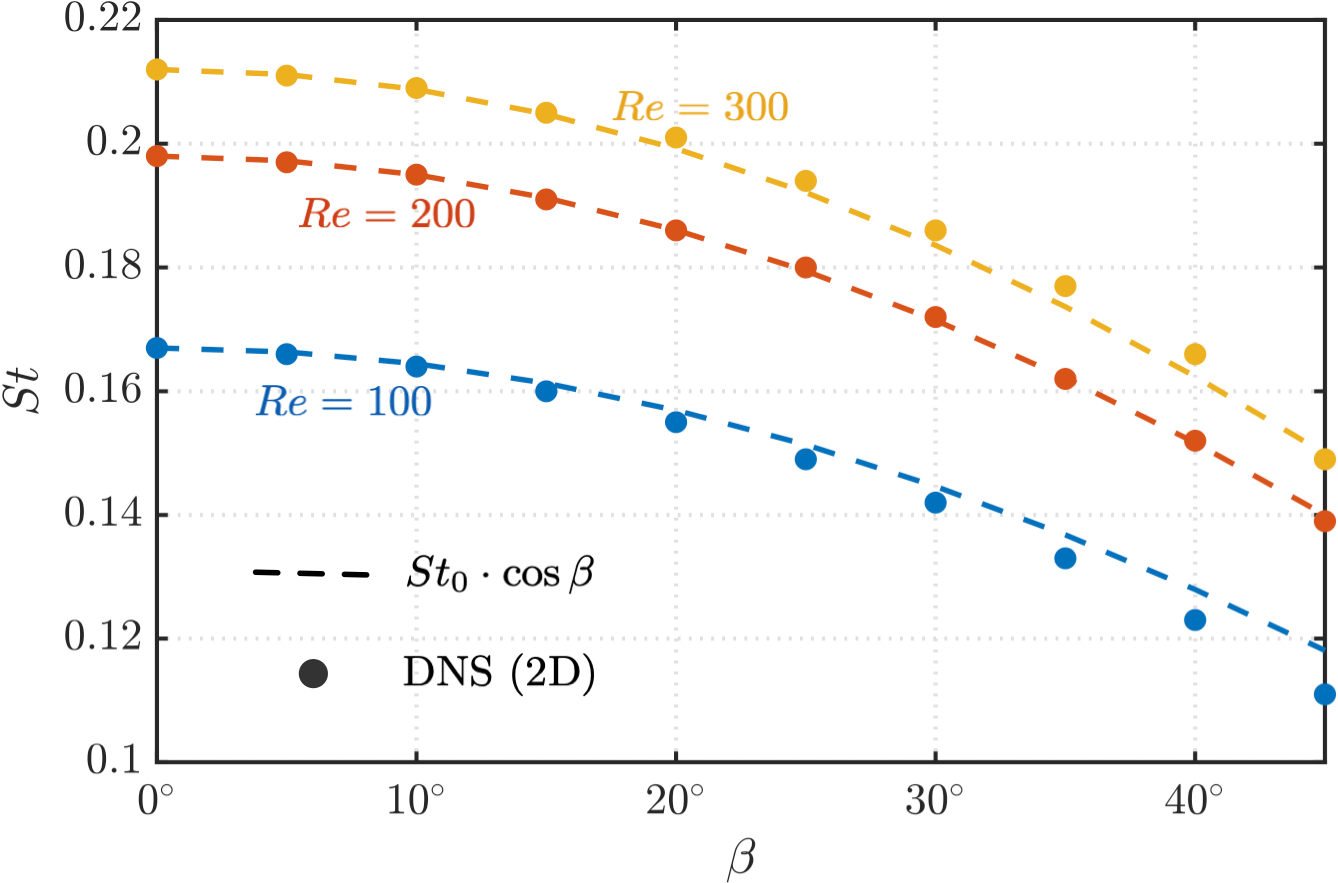}
\caption{Comparison of the Strouhal number between two-dimensional DNS and the cosine law.}
\label{fig:2DCosineRule}
\end{figure}

In the current study, the flow over inclined cylinder is simulated by shearing the computational domain with fixed incoming velocity. 
As a result, the velocity component that is normal to the cylinder axis ``sees" the cross-sectional shape of the cylinder as an ellipse.
This is different from some of the previous studies, which employed a setup where the velocity direction is altered to change the inclination angle, such that the velocity component that is normal to the cylinder axis still ``sees" a circular shape \citep{zhao2013three,zhang2018large}.
To assess the validity of the independence principle in the current setup, we carry out two-dimensional simulations (with the intrinsic secondary instabilities suppressed) of flow over inclined cylinder, and compare the results with the cosine law.
As shown in figure \ref{fig:2DCosineRule}, the results from the 2D DNS are in good agreement with those predicted by the independence principle, particularly at low inclination angles.
At higher $\beta$, the cosine law slightly overpredicts and underpredicts the shedding frequencies for $\Rey=100$ and 300, respectively.
We conclude that the independence principle can be used for estimating the shedding frequencies of inclined cylinders with the computational setup adopted in this study.

\section*{Acknowledgement}
KZ is grateful for the Office of Advanced Research Computing (OARC) at Rutgers, The State University of New Jersey for providing access to the Amarel cluster, on which some of the simulations were performed.
DZ thanks the support from Program for Intergovernmental International S\&T Cooperation Projects of Shanghai Municipality, China (grant no. 22160710200).
ZLH acknowledges the financial support from National Science Foundation of China (grant no. 52122110).

\bibliography{reference}
\bibliographystyle{jfm}
\end{document}